\documentclass[rsi,amsmath,amssymb]{revtex4-2}
\usepackage{graphicx}% Include figure files
\usepackage{dcolumn}% Align table columns on decimal point
\usepackage{bm}% bold math
\usepackage[utf8]{inputenc}
\usepackage[T1]{fontenc}
\usepackage{mathptmx}
\usepackage{upgreek}
\usepackage{units}
\usepackage{graphicx}
\usepackage{caption} 
\usepackage{subcaption}
\usepackage{comment}
\usepackage{wasysym}
\usepackage{textcomp}
\DeclareUnicodeCharacter{2212}{-}

\preprint{}
\begin{document}

\textit{\small Preprint: The following article has been submitted to \uppercase{{Review of Scientific Instruments}}}

\title{A compact low energy proton source}

\author{A. Weiser$^{a,b}$}
\author{A. Lanz$^{a,b}$}
\author{E. D. Hunter$^a$}
\author{M. C. Simon$^a$}
\author{E. Widmann$^a$}
\author{D. J. Murtagh$^a$}
\affiliation{$^a$Stefan-Meyer-Institute for Subatomic Physics, Austrian Academy of Sciences, Kegelgasse 27, 1030, Wien, Austria}
\affiliation{$^b$Vienna Doctoral School in Physics, University of Vienna, Vienna, Austria}

\date{\today}

\begin{abstract}
A low energy proton source for non-neutral plasma experiments was developed. Electrons from a hot filament ionize H$_2$ gas inside a geometrically compensated Penning trap to produce protons via dissociative ionization. A rotating wall electric field destabilizes the unwanted H$_2^+$ and H$_3^+$ generated in the process while concentrating protons at the center of the trap. The source produces bunches of protons with relatively low ion contamination (5.5 \% H$_2^+$ and 15.5 	\% H$_3^+$), with energy tunable from 35 to 300 eV.
\end{abstract}

\maketitle

\section{\label{sec:introduction}Introduction}
The ASACUSA Cusp experiment aims to measure the hyperfine splitting of antihydrogen in a magnetic field free region \cite{mohri_possible_2003,widmann_measurement_2004,Kolbinger2021,kuroda2014source}.  The Cusp experiment produces antihydrogen via the three-body recombination process i.e. $\textrm{e}^+ + \textrm{e}^+ + \overline{\textrm{p}} \rightarrow \textrm{e}^+ + \overline{\textrm{H}}$ in a nested Penning trap \cite{quin:93}. To produce a beam of atoms with the desired attributes (ground state, spin polarized, mean speed $v<1000\,\mathrm{m/s}$), the plasma properties of the antiprotons and positrons must be optimized. \par

Optimizing the mixing process is challenging and requires repeated experiments with positron and antiproton plasmas. Positrons are produced by a $^{22}$Na source and hence are available at any time. Trappable antiprotons are only available from CERN's Antiproton Decelerator (AD), which is not always online. For example, the entire complex was shut down for two years (2019-2021) while the LHC was upgraded. A low energy proton source allows ASACUSA to continue studying the mixing process during such periods by combining protons and electron plasma instead of antiprotons and positron plasma.\par
Such a source should preferably produce proton bunches comparable to those of the antiprotons as they enter the mixing trap ($\sim 10^5 \ \overline{\textnormal{p}}$ every $\unit[110]{s}$, $E_{kin} = \unit[100]{eV}$), have a high proton fraction, and a good vacuum rating of $ < \unit[10^{-6}]{mbar}$ so that the beamlines and traps are not contaminated by the operation of the source. \par
Proton sources using different methods of ion production have been developed for various applications. Radiofrequency sources such as electron cyclotron resonance (ECR) proton sources are commonly used in accelerator physics as they can create high energy ($\sim \unit[10]{keV}$), high current ($\sim \unit[10]{mA}$) proton beams with little H$_2^+$ and H$_3^+$ contamination \cite{Roychowdhury2015,Baumgarten2011}. 
Laser-driven proton sources can produce multi-MeV, ultra-short ($\sim$ ps) proton bunches and have applications in e.g. proton radiography and material research \cite{Bin2012,Barberio2018}. 
The use of such complex apparatuses is not necessary as such high currents and energies are not required for the planned operation.
Instead, a much simpler and more compact device using just electron bombardment of a H$_2$ gas is sufficient. 
Small ion sources have been developed for other particle traps \cite{Murbock2016} and are also commercially available (e.g.: SPECS, IQE 11/35) but they have not addressed the issue of active suppression of H$_2^+$ and H$_3^+$.
%%%%%%%%%%%%
The proton source discussed here uses electron impact ionization of H$_2$ gas and a Penning trap gas cell to produce a pulsed proton beam. 
In our case, H$_2^+$, and H$_3^+$ contaminants were reduced by applying a rotating wall (RW) electric field \cite{rotating-wall} using the azimuthally segmented central electrode of the gas cell. The proton source design and method of operation will be described in detail in Section~\ref{sec:proton-source}. Results from characterization measurements will be shown in Section~\ref{sec:experimental-results}.

\section{Proton Source \label{sec:proton-source}}
A schematic diagram of the low energy proton source is shown in Fig.~\ref{fig:eandb-gascell}(a). The source is constructed of 3 modules: \textbf{A.} electron gun, \textbf{B.} gas cell trap, \textbf{C.} ion extraction and steering.

\subsection{Electron Gun}
\label{sec:electrongunmodule}
The electron beam is formed using a tungsten filament from Kimball Physics (ES-020), biased to the same potential as the surrounding housing electrode ($\sim$ \unit[20]{V}). The anode ($\sim$ \unit[100]{V}) extracts and focuses electrons out of the gun through a $\unit[2]{mm}$ aperture into the gas cell. 
Varying the filament heater current from $\unit[1.6]{A}$/$\unit[0.9]{V}$ to $\unit[2.1]{A}$/$\unit[1.58]{V}$, electron emission currents between $\unit[2.1]{pA}$ and \unit[4.8]{\textmu A} were reproducibly generated. The emission current was determined by measuring the voltage drop over a resistor from the anode to ground. The resistors used in this measurement were chosen such that the anode potential did not change by more than $\unit[\sim 0.5]{V}$. The filament and housing were typically operated with a positive bias so that any grounded surface would repel the electrons.\par 

\subsection{Gas Cell Trap}
\label{sec:gascellmodule}
This module serves three purposes. First, it contains the H$_2$ gas which the electron beam ionizes. Second, it produces magnetic and electric fields similar to those of a Penning trap to confine the ions before pulsed ejection. Finally, it allows the application of the RW to drive out contaminant ions.\par

The cell is constructed from 6 electrodes: an entrance cap, an end cap, and a body that is azimuthally segmented into 4 petals, which can be individually biased. The entrance electrode has a $\unit[2]{mm}$ diameter, $\unit[4]{mm}$ long entrance tube. This allows the electron beam to enter and reduces the flow rate of H$_2$ into the region between the cell and gun. The exit electrode consists of a  $\unit[5]{mm}$ long tube with a larger diameter of $\unit[4]{mm}$. Although this allows more gas into the region between the cell and the ion extraction region, it also gives a larger solid angle for proton extraction. The central body is $\unit[28]{mm}$ long and is split into 4 electrodes. The flow rate of hydrogen is externally controlled by a simple needle valve.\par

\begin{figure}[h]
	\centering
	\includegraphics[width=0.9\textwidth]{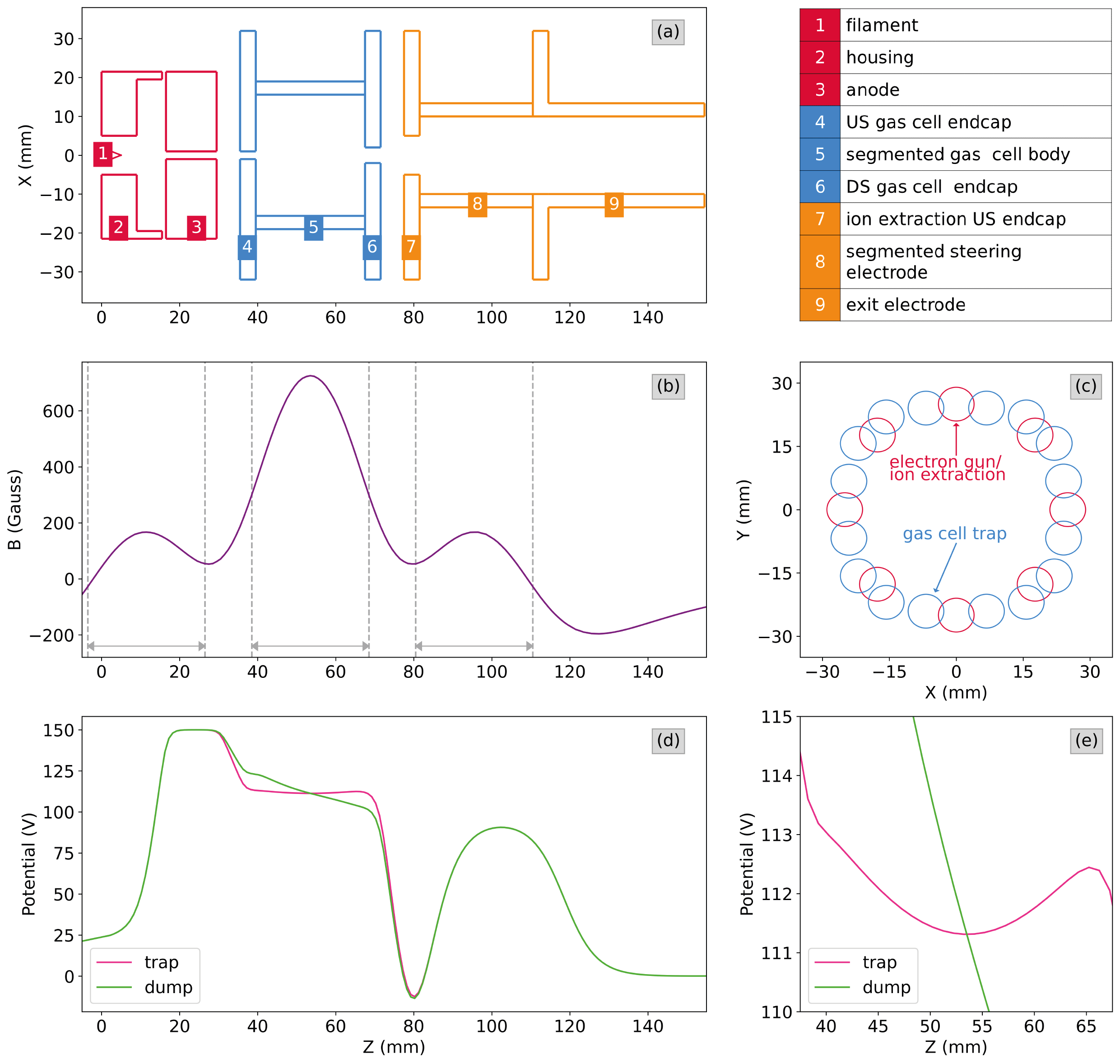}
	\caption{(a) Schematic diagram of a cut view through the electrodes of the proton source; red: electron gun, blue: gas cell trap, orange: ion extraction. (b) Z-component of the magnetic field strength (gauss) throughout the proton source; the dashed lines and arrows indicate the axial extent of the permanent magnets. (c) Azimuthal distribution of the magnets. They are evenly distributed ($r = $ \unit[25]{mm}) in rings of 8 around the electron gun and ion extraction module. The gas cell trap is surrounded by two rings of 8 with different radii ($r_1 = $ \unit[25]{mm}, $r_2 = $ \unit[27]{mm}) to allow space for the electrode contacts and gas input. (d) Electrical potentials used during trapping (pink line) and extracting (green line) from the trap (e) A closer view of the trapping/dumping potential.}
	\label{fig:eandb-gascell}
\end{figure}

The gas cell can be operated without applying a trapping potential. In this mode, the ions produced by the electron beam are continuously accelerated out of the cell by a constant linear ramp-type potential (see Fig.~\ref{fig:eandb-gascell}(c) green line). To produce a trap potential, the entrance and exit electrodes are biased higher than the central electrode (see Fig.~\ref{fig:eandb-gascell}(c) pink line). The trap is emptied by pulsing the Entrance (Exit) electrode up (down), again forming a ramp type potential, also allowing for time-of-flight spectroscopy. \par

Penning traps confine charged particles by superimposing a quadrupole electric field and a uniform magnetic field \cite{blau:10}. In the gas cell trap, the latter is provided by permanent magnets, which are a reasonable alternative when a fixed field value fulfills the requirements. In the present case, a total number of 32 neodymium (NdFeB) rod magnets (remanence $B_r$ = \unit[1.3]{T}, length = \unit[30]{mm}, diameter = \unit[8]{mm}) arranged as shown in Fig.~\ref{fig:eandb-gascell}(b-c) create a field with a strength of $\unit[73]{mT}$ in the center. The magnets around the electron gun and ion extraction module were added to prevent the axial magnetic field from changing direction between the modules. The z-component of the magnetic field, simulated in COMSOL Multiphysics 6.1, is shown in Fig.~\ref{fig:eandb-gascell}(b). The azimuthal asymmetry $\Delta B / B$ is $\sim 4.13\%$ at $r= \unit[5]{mm}$ off the trap axis. Producing such a field with normal conducting solenoid magnets as frequently used for Penning traps usually requires large magnets outside the vacuum vessel. The electrodes described above follow the design for geometrically compensated Penning traps \cite{GABRIELSE19841}. They provide a suitable potential to trap protons at the simulated magnetic field strength.
% COMSOL Multiphysics \textsuperscript \textregistered 6.1

The cross-section for the process $\textrm{e}^- + \textrm{H}_2 \rightarrow 2\textrm{e}^- + \textrm{H}^+_2$ is an order of magnitude larger than for the dissociative ionization process $\textrm{e}^- + \textrm{H}_2 \rightarrow 2\textrm{e}^- + \textrm{H} + \textrm{H}^+ $ \cite{Straub1996AbsolutePC}. Many of the H$_2^+$ formed will also interact in the gas cell producing H$_3^+$ via the process $\textrm{H}_2^+ + \textrm{H}_2 \rightarrow \textrm{H}_3^+ + \textrm{H}$ \cite{H3+}. The dissociation of molecular hydrogen by electron impact has been well studied \cite{dunn_dissociative_1963,crowe_dissociative_1973}. Dissociation proceeds either via the attractive ${^2 \Sigma _g^+}$ state which produces thermal to \unit[3]{eV} protons \cite{dunn_dissociative_1963} or the repulsive $^2 \Sigma _u^+$ state which produces protons with energies up to \unit[10]{eV} \cite{crowe_dissociative_1973}. 
The source is designed such that in trapping mode, the RW will stabilize the motion of protons while removing the unwanted H$_2^+$ and H$_3^+$ ions. Ions heavier than H$_3^+$ do not fulfill the confinement condition $ \left( \omega_c^2 > 2\omega_z^2, \textnormal{ where } \omega_c \textnormal{ is the cyclotron- and } \omega_z \textnormal{ the axial- frequency in the trap} \right)$, of this Penning trap and leave the trap in less than \unit[15]{\textmu s} (for a well depth of $\sim$ \unit[1]{V}).

\subsection{Extraction Module}
\label{sec:extraction-module}
It is important to guide the protons into the Cusp trap on axis. This is done by the extraction module which focuses and steers the proton beam. Ions exiting the gas cell are accelerated by the ion extraction electrode, then can be steered by a 4-fold segmented cylinder. Electrostatic lenses can be produced between the steering electrode and the exit electrode for focusing the beam as it is accelerated out of the source.   

\section{Characterization \label{sec:experimental-setup}\label{sec:experimental-results}}
The source was characterized using a position-sensitive MCP delay-line detector (DLD40) from Roentdek \cite{JAGUTZKI2002256}, installed in a second differentially pumped chamber downstream of the proton source. To protect the MCP from high-intensity ion beams, the filament was operated below nominal conditions. It was electrically heated by $\unit[1.725]{A}$/$\unit[1.07]{V}$ ($I_0 \sim \unit[84]{pA}$) instead of $\unit[2.5 - 2.8]{A}$.
The electron impact energy was varied by increasing or decreasing the gas cell potential with respect to the fixed potential of the electron gun filament. This also changes the energy of the ions extracted from the gas cell. For initial testing, the electron energy was set to $\unit[85]{eV}$, near the peak of the proton production cross-section \cite{Straub1996AbsolutePC}. This corresponds to an ion energy of $\unit[105]{eV}$. \par
The source was first tested in continuous ion extraction mode i.e. without applying a trapping potential in the gas cell (see Fig.~\ref{fig:eandb-gascell}(d)). Measurements were conducted to confirm the extraction module's ability to focus and move the ion beam before attempting to trap.  Through variation of the extraction ring potential relative to the gas cell potential the beam could either be focused or defocused at the detector, which was $\unit[\sim 46]{cm}$ away from the source (measured to the center of the gas cell). The application of different voltages on the split electrode steered the beam in the desired direction \cite{Weiser:2748624}. \par
\begin{figure}[h]
	\centering
	\includegraphics[width=0.6\textwidth]{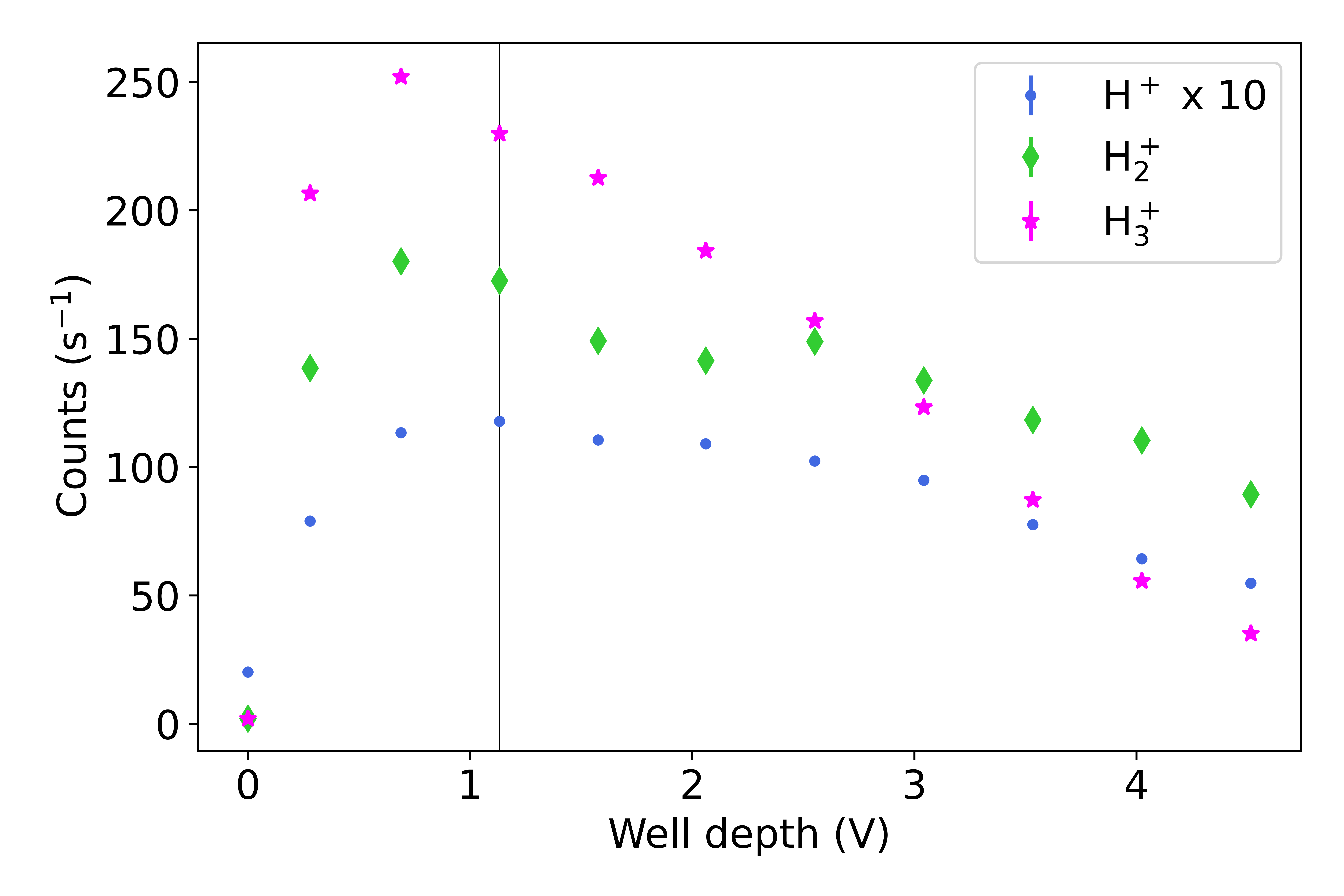}
	\caption{Number of extracted ions of each species per second depending on the well depth (extraction frequency = $\unit[1.25]{kHz}$, measurement time = \unit[600]{s}). The line shows the selected well depth for the operation of the source.}
	\label{fig:well-depth}
\end{figure}
In trapping mode, the optimal well depth was determined by maximizing the number of protons extracted from the source. The different ion species were identified using time-of-flight spectroscopy (see below). Figure~\ref{fig:well-depth} shows the number of ions of each species measured for different well depths. For this scan, the voltage on the entrance and exit electrodes was varied with all other potentials\textemdash and the overall energy\textemdash held constant. As the proton yield was the highest for a voltage difference of $\unit[3]{V}$ between the gas cell endcaps and the ring electrode, which corresponds to an on-axis well depth of \unit{1.13}{V}, this well depth was selected for the operation of the source. \par
\begin{figure}[h] 
	\centering
	\includegraphics[width=\textwidth]{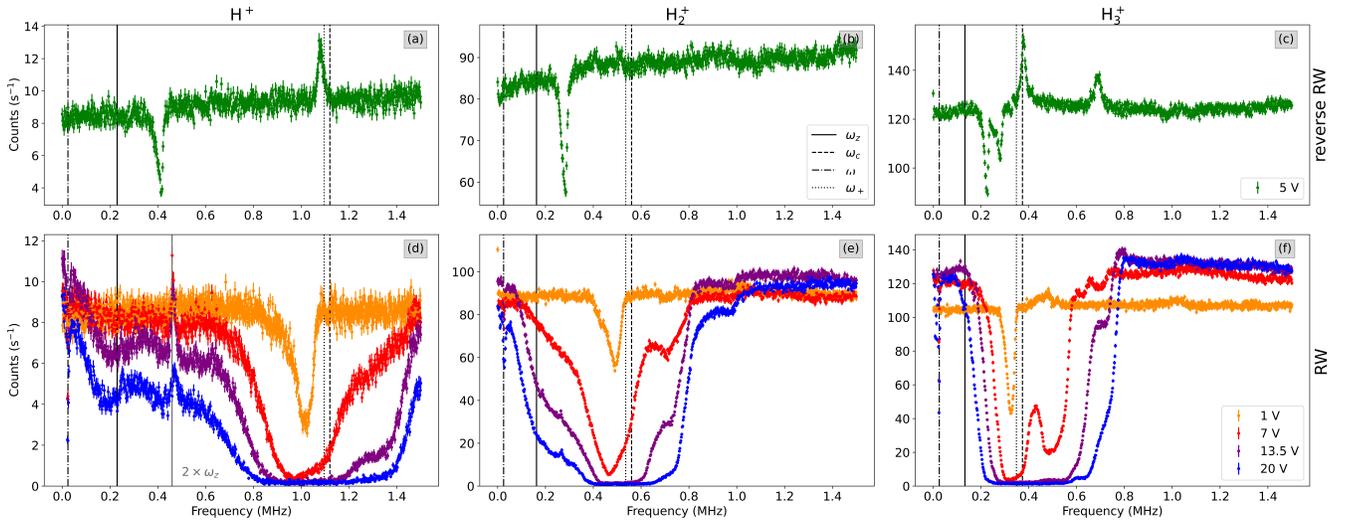}
	\caption{Rotating wall frequency scans. The figures compare protons, H$_2^+$, and H$_3^+$ counts for different RW amplitudes. Figures at the top are counter-rotating. Also indicated on these figures in lines are the magnetron (dashed-dotted), axial bounce (solid), cyclotron (dashed), and modified cyclotron frequency (dotted).}
	\label{fig:rotwall-reversed-freq-1}
\end{figure}
\begin{figure}[h] 
	\centering
	\includegraphics[width=0.7\textwidth]{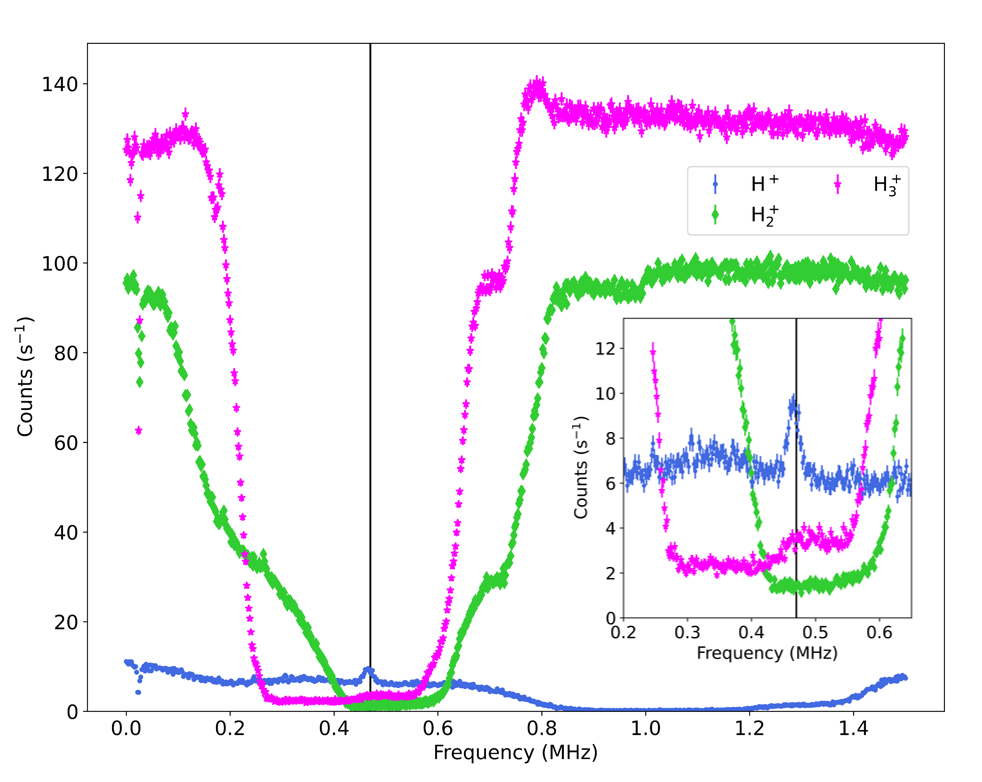}
	\caption{Comparison of proton (blue), H$_2^+$ (green) and H$_3^+$ counts (magenta) for a fixed amplitude of \unit[13.5]{V} (i.e. purple data of Fig.~\ref{fig:rotwall-reversed-freq-1}~(d-f)). The solid black line shows the selected frequency of \unit[0.47]{MHz}.}
	\label{fig:rotwall-reversed-freq-2}
\end{figure}

The frequency of the RW electric field was scanned, for different RW amplitudes, monitoring the rate of proton, H$_2^+$, and H$_3^+$ counts at the detector. Figure~\ref{fig:rotwall-reversed-freq-1}(d,e,f) shows the count rate obtained for each species as a function of RW frequency. Two Mini Circuits power splitters (ZSCJ-2-2+) were used to split the RW perturbations, reducing their amplitude in the process. The loss of the hereinafter-mentioned voltages can be found in the corresponding datasheet. The scans shown here were conducted at RW amplitudes of \unit[1]{V} (orange points), \unit[7]{V} (red points), \unit[13.5]{V} (purple points) and \unit[20]{V} (blue points). In all cases, it appears that the application of a frequency close to the magnetron frequency ($\omega_-$) drove out each species (see also Fig.~\ref{fig:rotwall-reversed-freq-2}). In the case of protons, the application of a frequency close to twice the axial bounce frequency produced an increase in the observed count rate.
\par 
The main effect observable was the substantial loss of counts due to the dipolar excitation at the (mass-dependent) reduced cyclotron frequency. The effect broadened with the increase of RW amplitude. This cleaning of ion species via deliberate radial ejection has been used in different Penning trap setups such as ISOLTRAP \cite{Blaum_2004} and JYFLTRAP \cite{KOLHINEN2004776}.   
Consider Fig.~\ref{fig:rotwall-reversed-freq-2} which compares the count rates for protons, H$_2^+$, and H$_3^+$ for the \unit[13.5]{V} scan: by selecting a frequency close to twice the axial bounce frequency for protons (\unit[0.47]{MHz}), the H$^+$ count rate was preserved, whereas in the case of H$_2^+$, and H$_3^+$, the rates were greatly reduced by an apparent coupling to the reduced cyclotron motion.
For a counter-rotating field (Fig.~\ref{fig:rotwall-reversed-freq-1}(a-c)), an increase in protons and H$_3^+$ close to their cyclotron frequency was observed, whereas no such effect could be seen for H$_2^+$ ions.
Furthermore, the counter-rotating field removed every species at roughly twice the (species-dependent) axial bounce frequency. However, this rotating field direction was not chosen for further operation, as the co-rotating field both preserves the proton yield and simultaneously drives out H$_2^+$, and H$_3^+$.

\par
\begin{figure}[h] 
	\centering
	\includegraphics[width=0.7\textwidth]{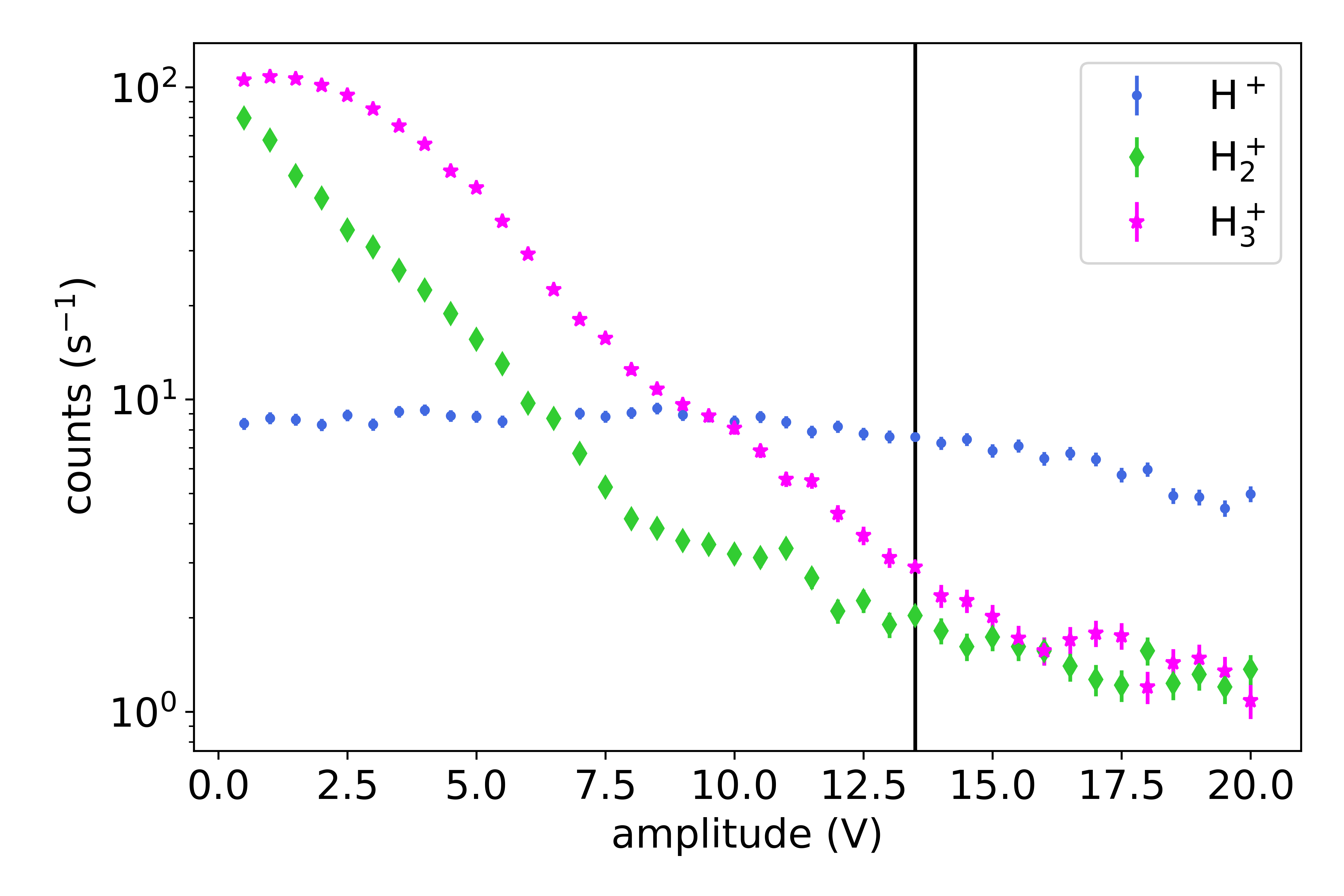}
	\caption{Amplitude scan at a RW frequency of $\unit[0.47]{MHz}$. The solid black line shows the selected amplitude of \unit[13.5]{V}.}
	\label{fig:rotwall-reversed-ampl-1}
\end{figure}

The optimal RW amplitude was determined by scanning this parameter (i.e. the oscillating voltage $V_{pp}$ applied to the electrodes) with the frequency fixed at \unit[0.47]{MHz}. Figure~\ref{fig:rotwall-reversed-ampl-1} shows that the number of protons was reasonably constant across the range investigated (as may be expected from Fig.~\ref{fig:rotwall-reversed-freq-1}(d)), and for $V_{pp} \geq 10$ more protons were extracted than H$_2^+$ or H$_3^+$. 

\begin{figure}[h] 
	\centering
	\includegraphics[width=0.7\textwidth]{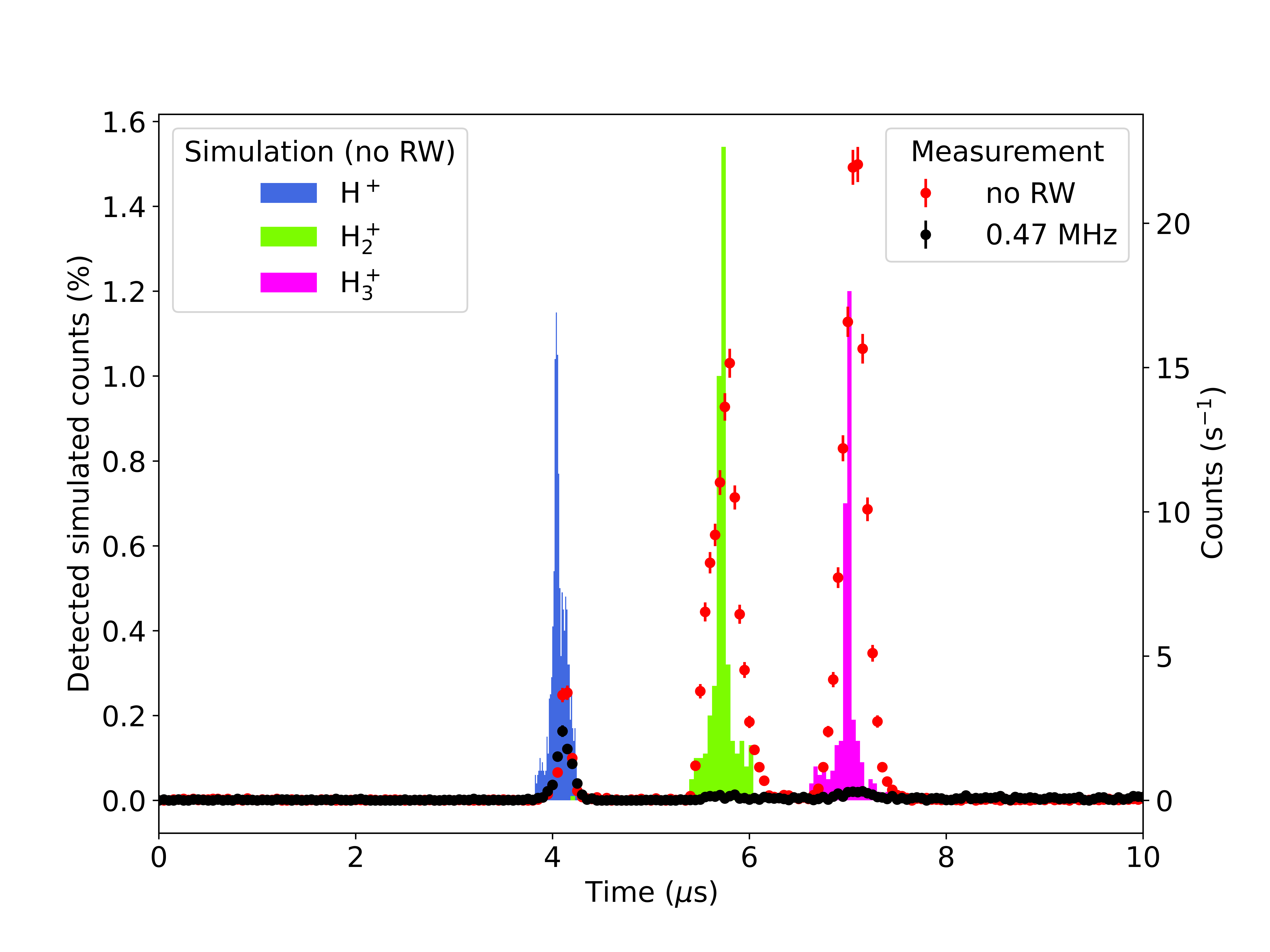}
	\caption{Measured time of flight spectrum with and without the RW (f = \unit[0.47]{MHz}, $V_{pp}$ = \unit[13.5]{V}), and simulation of the extraction out of the gas cell}
	\label{fig:rotating-wall-tof}
\end{figure}

The largest difference between the proton and other ion count rates was achieved at about $\unit[13.5]{V}$. At this point $\unit[(8.67 \pm 0.38)]{s^{-1}}$ protons, $\unit[(1.38 \pm 0.15)]{s^{-1}}$ H$_2^+$ and $\unit[(3.45 \pm 0.24)]{s^{-1}}$ H$_3^+$ were measured. 
\\
A time-of-flight spectrum at these optimum values is shown in Fig.~\ref{fig:rotating-wall-tof} in black. The proton peak is well-defined at \unit[4.1]{\textmu s}. The H$_2^+$ and H$_3^+$ peaks at \unit[5.8]{\textmu s}  and \unit[7.1]{\textmu s} are visible but smaller than the proton peak.
This figure also shows the time-of-flight spectrum without the application of the RW, in red.
The extraction of 10000 ions of each species was simulated in SIMION 8.1, and the time-of-flight to the detector was recorded. These results are included in Fig.~\ref{fig:rotating-wall-tof} for comparison. In reality, due to the cross-sections \cite{Straub1996AbsolutePC} we expect approximately 10x less H$^+$ than H$_2^+$ in the trap. At the voltage chosen for the MCP operation, the detection efficiencies for all species are the same \cite{PEKO2000597}. The simulations showed a total arrival efficiency at the detector (combination of the ion extraction efficiency from the source and the transport efficiency to the detector) of $\approx 11.6\% $ for protons, $\approx 4.3\% $ for H$_2^+$ and $\approx 3.2\% $ for H$_3^+$. 
While the extraction efficiency did not depend greatly on the axial position of the particles in the gas cell, on-axis particles were favored with the extraction efficiency dropping to half of its maximum at $r = \unit[2]{mm}$. Any further interactions between the ions and H$_2$ or residual gas on the way to the detector were not taken into account. 

\begin{figure}[h] 
	\centering
	\includegraphics[width=0.7\textwidth]{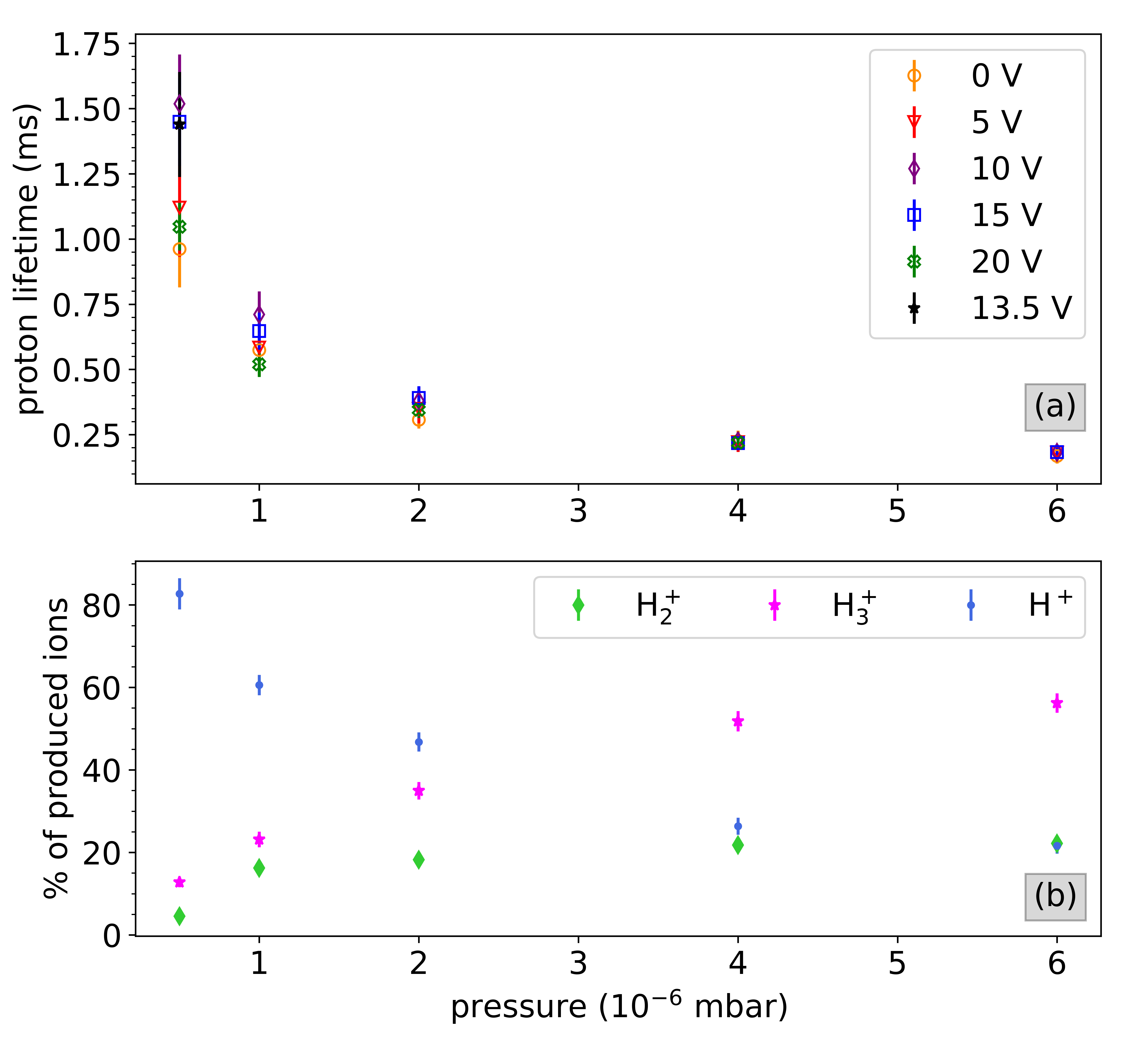}
	\caption{a) Lifetime of protons in the gas cell trap as a function of vacuum chamber pressure and RW amplitude at a RW frequency of $\unit[0.47]{MHz}$; b) Percentages of extracted ion species for different vacuum chamber pressures at RW settings (f = \unit[0.47]{MHz}, $V_{pp}$ = \unit[13.5]{V})}
	\label{fig:rotwall-reversed-lifetime}
\end{figure}

The lifetime of the protons in the trap was estimated by measuring the number of protons produced for different trapping times. The trapping time was varied by changing the extraction frequency while keeping the extraction pulse width constant, i.e. the trap was filled for different time spans.
An exponential saturation curve of the form $N \sim a \cdot exp[-t/\tau] + b$  was fitted, where $\tau$ is the proton lifetime in the trap \cite{Weiser:2748624}.
Figure~\ref{fig:rotwall-reversed-lifetime}(a) shows that the lifetime decreased with increasing gas pressure, indicating that a higher collision rate was detrimental to stable trapping.
The lifetime also depended on the RW amplitude, peaking around $\unit[10-15]{V}$. 
The stabilization of protons in the trap due to the application of twice the axial bounce frequency improves with increasing amplitudes, however, so does the power broadening of the radial ejection at the reduced cyclotron frequency. The reduction in lifetime at higher amplitudes could therefore be attributed to the second effect becoming dominant.
 
Figure~\ref{fig:rotwall-reversed-lifetime}(b) shows the fraction of the extracted ions species for different pressures, using the optimal RW settings (f = \unit[0.47]{MHz}, $V_{pp}$ = \unit[13.5]{V}).
The unfavorable reduction of the H$^+$-ratio at higher pressures might be attributed to the larger number of secondary interactions undergone by protons in the extraction module. \par

The pressure reported in Fig.~\ref{fig:rotwall-reversed-lifetime} was measured in the proton source vacuum chamber and does not correspond to the pressure in the gas cell. A pressure simulation of the whole system was therefore conducted in Molflow+ \cite{Kersevan:2019vme} using 3D renderings of the source and a to-scale representation of the vacuum system. 
Pumping speeds were set according to the values given by pump manufacturers and the H$_2$ flow-rate was varied.
We obtain that the pressures in the gas cell are $\sim 120$ times larger than the pressure measured at the pressure gauge. For the electron gun and ion extraction module, they are both higher by about a factor of 2.
Most of the work described here was conducted at $\unit[1 \times 10^{-6}]{mbar}$. This would correspond to a gas cell pressure of $\sim \unit[1.2 \times 10^{-4}]{mbar}$. \par

\section{Summary}
A simple low energy proton source, based on electron impact ionization of H$_2$ in a Penning trap gas cell, was constructed and characterized. It will be used by the ASACUSA Cusp experiment for comparative matter experiments especially when antiprotons are unavailable. As expected, in agreement with the H$_2$ ionization cross-section, the number of protons produced is approximately one order of magnitude lower than the H$_2^+$ and H$_3^+$ ions. A rotating wall electric field was applied to the gas cell to preserve the number of protons and reduce the number of H$_2^+$ and H$_3^+$ ions. This technique was successfully applied to reduce the number of background ions by an order of magnitude whilst maintaining the number of protons.\par
Energy tunable pulses of $\sim 9$ protons/s (extraction frequency = 500 Hz, vacuum chamber pressure = $\unit[0.5 \times 10^{-6}]{mbar}$)  were produced for the low electron intensity of $\sim \unit[84]{pA}$. These pulses have minimal contamination of $\sim 5.5\%$ H$_2^+$ and $\sim 15.5\%$ H$_3^+$. 
Although our single-particle detector does not allow for high-intensity rate measurements, our results would extrapolate to $\sim10^5$ protons/s when the electron emission is raised to $\sim$ \unit[4.8]{\textmu A}.
The proton source has now been installed into the ASACUSA apparatus and allows proton plasma experiments whenever antiprotons are not available. Furthermore, it may also be used to test new equipment integrated into the $\overline{\textnormal{p}}$-beamline.

\section*{Acknowledgements}
We would like to thank Doris Pristauz-Telsnigg, Leopold Stohwasser, Mark Pruckner, and Herbert Schneider for their expert technical assistance. This work was supported by the Austrian Science Fund (FWF) P32468 and W1252-N27.
\section*{Author Declarations}
\subsection*{Conflict of Interest}
The authors have no conflicts to disclose.
\subsection*{Author Contributions}
\textbf{A.~Weiser:} Formal analysis (lead); investigation (lead); visualization (lead); writing - original draft (lead); software(equal); validation (equal). \textbf{A.~Lanz:}  Writing - reviewing and editing (equal); investigation (equal); software(equal); validation (equal). \textbf{E.~D.~Hunter:} Methodology (equal); validation (equal); writing - reviewing and editing (equal). \textbf{M.~C.~Simon:} Conceptualization (lead); funding acquisition (equal); resources (equal); writing – reviewing and editing (equal); supervision (supporting). \textbf{E.~Widmann:} Funding acquisition (supporting); supervision (supporting); writing - reviewing and editing (supporting). \textbf{D.~J,~Murtagh:} Conceptualization(equal); Funding acquisition(lead); methodology(equal); project administration (lead); software (supporting); supervision (lead); validation (equal); visualisation (supporting); writing - reviewing and editing (equal).

\section*{Data Availability Statement}
The data that support the findings of this study are available from the corresponding author upon reasonable request.

\bibliography{bibliography}% Produces the bibliography via BibTeX.
\end{document}